\documentclass[twocolumn,twoside,slac_two]{revtex4}
\usepackage{graphicx}
\usepackage{fancyhdr}
\usepackage{hyperref}
\hypersetup{
    colorlinks=true,
    urlcolor=magenta,
}
\newcommand*{\xmax}{$X_\mathrm{max}$}
\newcommand*{\mxmax}{X_\mathrm{max}}
\newcommand*{\fig}{{FIG.}}
\newcommand*{\gcm}{$\text{g}/\mathrm{cm}^2$}

\begin{document}

\title{Radio detection of cosmic rays with the Auger Engineering Radio Array}

%

\author{Florian Gat\'e}
\affiliation{Subatech, Universit{\'e} de Nantes, \'Ecole des Mines de Nantes, CNRS/IN2P3, Nantes, France}
\author{for the Pierre Auger Collaboration}
\affiliation{For the complete author list, see \url{http://www.auger.org/archive/authors_2016_09.html}}

\begin{abstract}
The very low statistics of cosmic rays above the knee region make their study possible only through the detection of the extensive air showers (EAS) produced by their interaction with the constituents of the atmosphere. The Pierre Auger Observatory located in Argentina is the largest high energy cosmic-ray detection array in the world, composed of fluorescence telescopes, particle detectors on the ground and radio antennas. The Auger Engineering Radio Array (AERA) is composed of 153 autonomous radio stations that sample the radio emission of the extensive air showers in the 30 MHz to 80 MHz frequency range. It covers a surface of 17 km$^2$, has a 2$\pi$ sensitivity to arrival directions of ultra-high energy cosmic rays (UHECR) and provides a duty cycle close to 100\%. The electric field emitted by the secondary particles of an air shower is highly correlated to the primary cosmic ray characteristics like energy and mass, and the emission mechanisms are meanwhile well understood. In this contribution, recent progress on the reconstruction of the mass composition and energy measurements with AERA will be presented.
\end{abstract}

\maketitle

\thispagestyle{fancy}


\section{Introduction}
The Pierre Auger Observatory aims at solving the mysteries of the origin of the ultra-high energy cosmic rays (UHECR). Lots of progress have been done in the last 10 years: a clear cutoff is observed in the energy spectrum around $4\times 10^{19}$~eV~\cite{spectrumpao2010,Abbasi:2007sv,AbuZayyad:2012ru}, as well as a trend towards a heavier composition at the highest energies, the proton-air cross section is measured at $10^{18}$~eV, exotic scenarii are disfavored as no photon- or neutrino-like showers were detected... For a recent review of the Auger results, see~\cite{Ghia:2015kfz}. The Pierre Auger Collaboration will upgrade its instruments to increase the mass-discrimination capabilities. One way to achieve this goal is to exploit the radio signal emitted by air showers while developing in the atmosphere.

\section{The AERA experiment}

The Pierre Auger Observatory~\cite{ThePierreAuger:2015rma}, the largest UHECR observatory in the world, hosts a large surface detector (SD) of 1660 Cherenkov tanks spread over 3000~km$^2$, a fluorescence detector (FD) of 27 telescopes at 4 sites around the SD, the AMIGA instrument with buried scintillators for the measurement of the muonic component of air showers, and the Auger Engineering Radio Array (AERA).

\begin{figure}
\includegraphics[width=80mm]{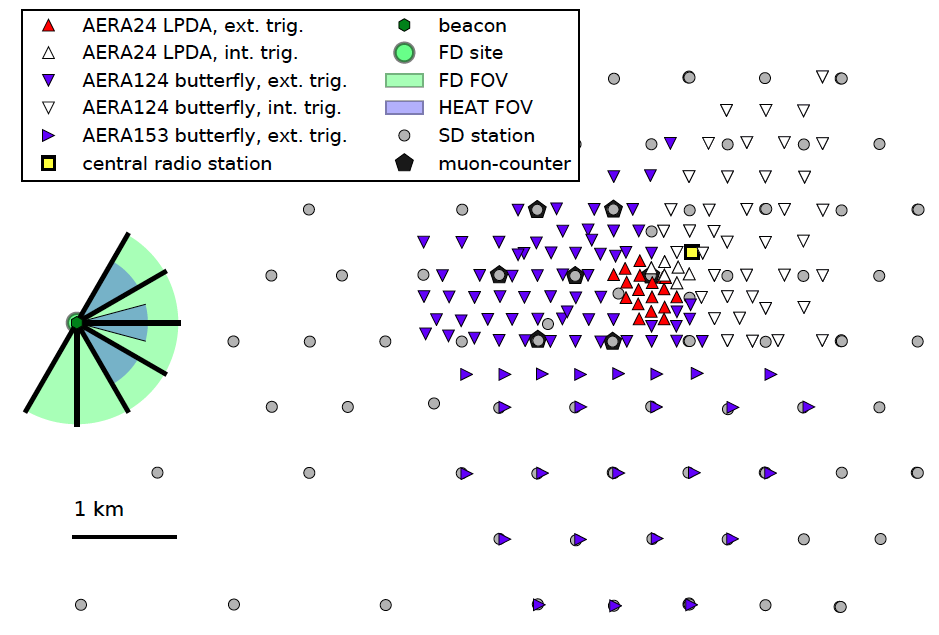}
\caption{Map of the AERA antennas (triangles) together with the other instruments deployed at the Pierre Auger Observatory (water
Cherenkov tanks, fluorescence telescopes and AMIGA muon counters).}
\label{aeramap}
\end{figure}

AERA consists today of 153 autonomous radio stations that detect the very fast (10-30~ns) transient radio signal emitted by the secondary electrons and positrons of air showers. The electric field is detected by antennas in the band 30-80~MHz (log-periodic dipoles~--- LPD --- in the case of the first 24 stations in 2011 and butterfly dipoles afterwards) and sampled by ADCs running at 180~MHz or 200~MHz depending on the electronics in use. \fig~\ref{aeramap} shows the current setup.
The stations are powered with solar panels and are used in both self-trigger and external-trigger mode. External triggers come from the SD and the FD. We present here the results obtained using the external-trigger data. They are calibrated in time by the usage of GPS receivers together with a beacon emission that allow to correct for clock drifts. We also use the radio transients emitted by airplanes. The overall timing accuracy is better than 2~ns~\cite{Aab:2015omo}.

The amplitude calibration has been studied in detail with the LPD antennas using a known calibrated source carried by weather balloons and, more recently, a remotely piloted drone. The final amplitude accuracy is 9.3\% accounting for all kinds of systematic uncertainties~\cite{krausevienna}. This calibration procedure will be applied on the butterfly antennas.

\section{Electric-field emission mechanisms}

Secondary charged particles created during air-shower development emit coherent radiation that extends from frequencies of some kHz up to some hundreds of MHz. The emission extends up to the GHz domain but it is incoherent, less interesting and abandoned today.
Electrons and positrons experience multiple diffusion, which is a random process, and the Lorentz force due to the geomagnetic field which is systematic. This results in a net current $\mathbf{j}$ perpendicular to the shower axis and leads to a net electric field. This is called the geomagnetic effect and is the dominant source of radio waves from air showers. This electric field is linearly polarized in the direction of $\mathbf{a}\times\mathbf{B}$ where $\mathbf{a}$ is the shower axis direction and $\mathbf{B}$ is the geomagnetic field direction. An observer detects the same polarization, independently of its position with respect to the shower axis.

The second mechanism of coherent emission is the excess of electrons over positrons. Indeed, positrons annihilate very quickly in air and, in addition to the shower electrons, more electrons are extracted from the medium through Compton, Bhabha and Moeller diffusions. This excess of electrons leads to a net and coherent electric field which is radially polarized with respect to the shower axis. This means that the corresponding polarization depends on the observer position with respect to the shower axis. Its amplitude is one order of magnitude smaller than the geomagnetic electric field.

Both mechanisms have been detected in the data of various experiments ~\cite{marinicrc2011,radioemissionprd,belletoile:hal-01138851}, in particular thanks to the strong progress made in the simulation codes.

These two electric fields interfere and the final electric field has a complex structure that cannot be described by a simple 1D lateral distribution function (LDF). We can use for instance a 2D function taking into account both mechanisms~\cite{Nelles:2014gma}, based on the code CoREAS~\cite{coreas2013} as a model of ground distribution of the electric field. This method has been used in AERA to extract the primary energy from the radio signal. We also used the code SELFAS~\cite{selfas2011} to construct the 2D ground distribution of the electric field in order to extract the shower \xmax\ (see section~\ref{xmax}).

\section{Primary cosmic ray characteristics from the radio signal}

The electric field emitted by air showers is now well understood. We can therefore use the simulation to correlate the data and the model to extract the primary cosmic-ray characteristics: its primary energy and its nature through the \xmax\ of the shower it created.

\subsection{Energy}

\begin{figure}[!ht]
\includegraphics[width=65mm]{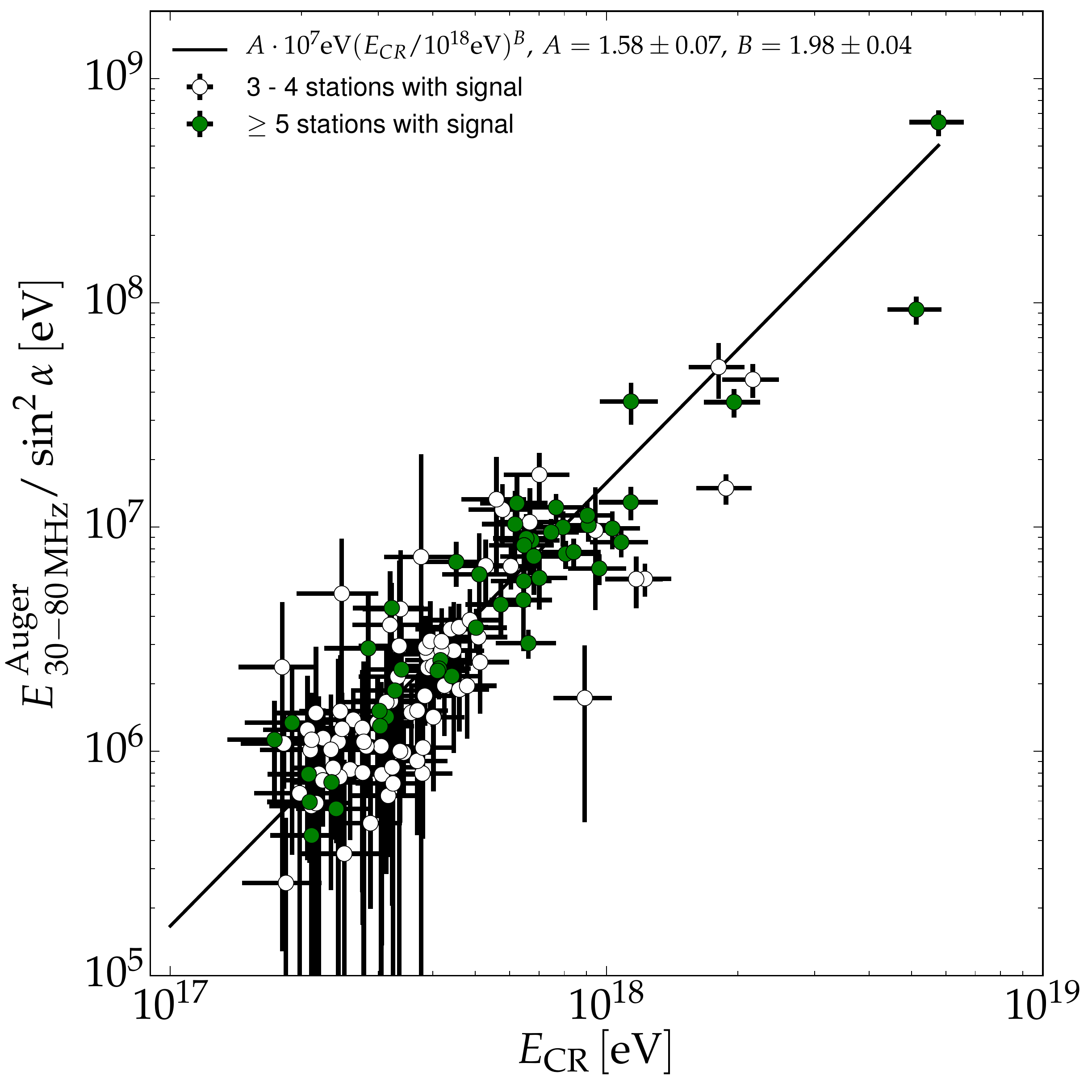}
\caption{Corrected radio energy as a function of the primary energy.}
\label{fig:nrj}
\end{figure}

A strong correlation has been established by AERA between the shower energy measured by the FD or the SD and a radio observable, having units of energy. This observable can be interpreted as the fraction of the primary energy that is radiated in the AERA frequency band 30-80~MHz. The data recorded by the radio stations are first corrected for the antenna and electronics responses~\cite{Abreu:2012pi}. After this step, we obtain the three components of the electric field as a function of time (together with the corresponding Poynting vector) for each station participating in an event. We integrate this Poynting vector over time to get the energy fluence (energy per unit area) for each station. This means that we have the energy fluence at each location of a station, i.e. the measured 2D ground distribution of the energy fluence. We fit these points with the 2D LDF. Once the best 2D LDF is obtained, we integrate it over ground coordinates to get the energy radiated in the 30-80~MHz band~\cite{Aab:2015vta,Aab:2016eeq}. We rescale the obtained radiated energy $E^\mathrm{Auger}_{30-80~\mathrm{MHz}}$ by the factor $1/\sin^2\alpha$ to take into account the angular distance $\alpha$ from the geomagnetic field. 
The corrected radiated energy is quadratically correlated to the primary energy $E_\mathrm{CR}$, as expected
in the case of a coherent mechanism. The correlation is shown in \fig~\ref{fig:nrj}.
From the measurement of the radiated energy in the 30-80~MHz band, we can invert the formula to extract the primary energy:
\begin{displaymath}
\frac{E_\mathrm{CR}}{10^{18}~\mathrm{eV}} = \left(\frac{E^\mathrm{Auger}_{30-80~\mathrm{MHz}}}{10^7~\mathrm{eV}}\cdot\frac{1}{A \sin^2\alpha}\right)^{1/B}
\end{displaymath}
with $A=1.58\,\pm\, 0.07$ and $B=1.98\,\pm\, 0.04$. For a primary cosmic-ray energy of  $10^{18}$~eV and an angular separation from the geomagnetic field of $90^\circ$, we have a radiated energy in 30-80~MHz of $15.8\pm 0.7~\text{(stat)}\pm 6.7~\text{(sys)}$~MeV.

\subsection{\xmax\ estimation}\label{xmax}

In this section, we explain how we perform the measurement of the shower \xmax\ using directly the simulation (here with SELFAS), not using the 2D LDF as in the previous section. The basic idea is to exploit the fact that the shape of the ground distribution of the electric field depends strongly on the distance between the shower core and the point of maximum emission on the shower axis. This distance in turn strongly depends on the mass of the primary cosmic ray.

\begin{figure}[!ht]
\includegraphics[width=80mm]{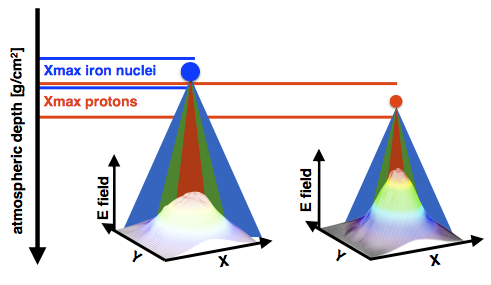}
\caption{Dependence of the footprint of the radio emission on the mass of the primary. Two LDF are simulated with SELFAS: iron nucleus (left) and proton (right). The vertical axis represents the atmospheric depth and the horizontal lines account for the typical \xmax\ distributions for proton-induced showers and iron-induced showers.}
\label{fig:topology}
\end{figure}

At fixed energy and zenith angle, light nuclei are more likely to interact at lower altitudes than heavier nuclei. The electric field emission is highly beamed towards the direction of propagation of the shower producing statistically a narrower LDF in the case of light nuclei compared to heavier nuclei. Thus, the topology of the electric field at the ground level is highly correlated to the mass of the primary, as depicted in \fig~\ref{fig:topology}.

\begin{figure}[!ht]
\includegraphics[width=80mm]{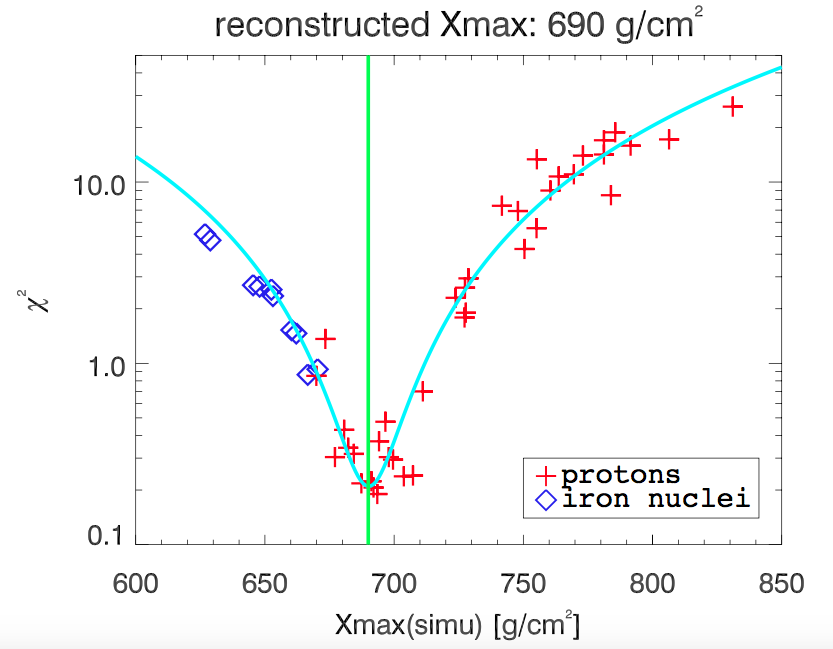}
\caption{Chi-square for the fit of a detected LDF at AERA and the simulated LDFs as a function of their respective \xmax\ depths. The values are fitted by a square function represented by the curve. The \xmax\ depth reconstructed by the radio method for this event is given by the minimum of the quadratic function (i.e. the value that gives the best agreement), highlighted on the plot by the vertical line.}
\label{fig:chi2xmax}
\end{figure}

The reconstruction method is based on a comparison of the amplitude of the detected electric field to its simulation. To reconstruct one detected shower, we use a set of simulated events having the measured arrival direction, using an arbitrary energy of $10^{18}$~eV and with realistic \xmax\ depths for showers initiated by protons and iron nuclei at this energy. The total set is composed of a higher fraction of protons to account for the larger width of the distribution of the first interaction depths for the light nuclei. The center of the simulated LDFs is moved to several positions
on the surface of the AERA array. For each tested position, the agreement between the data and the simulation is tested with a $\chi^2$ test. This first step permits us to reconstruct the position of the shower core. During this step the simulated amplitudes are multiplied by a scaling factor defined as the mean amplitude ratio between the data and the simulation. As the electric field amplitude depends linearly on the primary energy, the value of the scaling factor at the core position allows us to reconstruct the primary energy. The $\chi^2$ values obtained with each simulated LDF at the reconstructed core position allows the reconstruction of the detected shower \xmax\ as the minimum of the function $\chi^2 = f(\mxmax)$ as shown in \fig~\ref{fig:chi2xmax}.

\begin{figure}
\includegraphics[width=80mm]{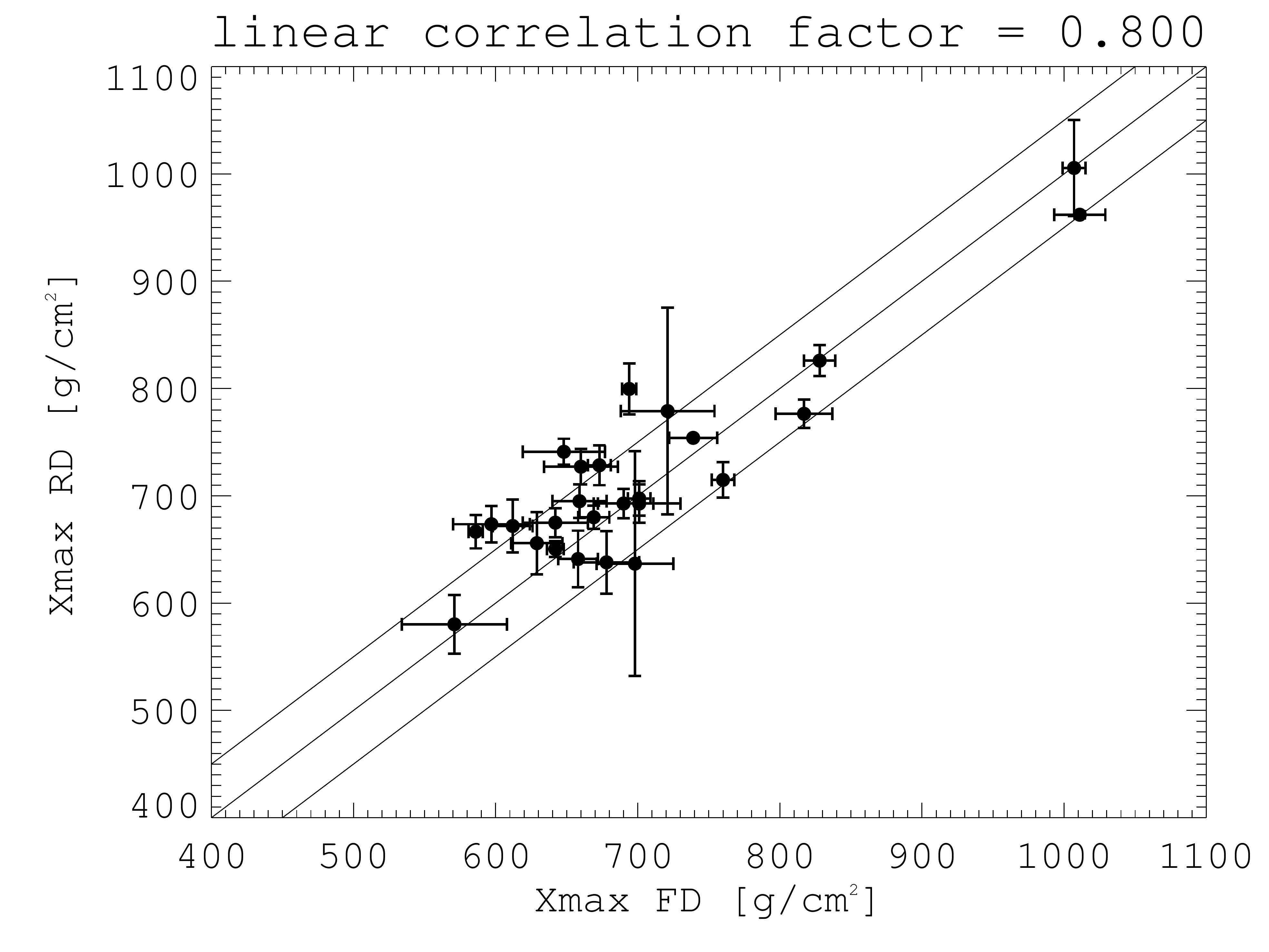}
\caption{Reconstructed \xmax\ with the radio method as a function of the FD measurements, the middle line accounts for a one-to-one correlation and the others account for a deviation of $\pm 50$~\gcm.}
\label{fig:xmax}
\end{figure}

The method is applied on a high quality set of multi-hybrid (radio, FD and SD) showers passing the official FD quality cuts; we also require that at least 5 AERA radio stations participate in the event and that the shower has a zenith angle smaller than $55^\circ$.
The correlation between the \xmax\ reconstructed by the radio method and the FD signal is shown in \fig~\ref{fig:xmax}.

The reconstructed values are in good agreement with the FD measurements and the mean deviation is compatible with zero with a dispersion of $25$~\gcm. It is important to note that these results are obtained using a realistic simulation of the atmosphere in SELFAS. To do so, we used the Global Data Assimilation System (GDAS)~\cite{gdas}. It gives information (pressure, temperature and air humidity) that allows the computation of the air density as a function of the altitude on a 3~hour basis, in the neighbourhood of the AERA site. For each detected event, the data sets necessary to perform the radio reconstruction are simulated using the air density and refractivity profiles matching the actual experimental conditions at the moment of the detection. The use of a stationary and fixed atmosphere description such as the US Standard model leads to a systematic shift of $17$~\gcm\ between the FD measurements and the radio method together with a much larger dispersion. Thus the simulation of the atmosphere matching the experimental conditions is now mandatory to reach the FD precision. Using this method with a realistic description of the atmosphere, we found the same shower core than the one obtained with the SD and FD: the average difference is smaller than $3$~m and the dispersion is smaller than $10$~m. The energy is also well reconstructed as the mean relative difference and dispersion to the SD reconstruction are $3$\% and $25$\%, respectively.

\section{Conclusion}

The latest results provided by AERA show that the radio signal contains the information needed to reconstruct all characteristics of the primary cosmic ray. The primary energy is estimated using the measured radiated energy in the  30-80~MHz band in use in AERA, from a 2D LDF based on CoREAS simulations. The estimation is unbiased and the resolution is of the order of $17\%$. This radiated energy estimator is the same at any experiment site as it is normalized to the geomagnetic field strength. It is hardly dependent on environmental conditions, contrarily to the fluorescence method because the atmosphere is transparent to radio waves in our frequency range of interest. Finally, this is a reliable method as it is based on the classical emission of the electromagnetic part of the shower, which is well understood.

A pure radio method (without the need of other detectors), using only the electric field measurements, has been developed using SELFAS simulations. It has been tested on multi-hybrid (radio, SD, FD) events detected in Auger. The energy and core position are accurately reconstructed and the shower \xmax\ is in excellent agreement with the FD estimation. This result is obtained when using a realistic description of the atmosphere provided by the GDAS. This is not the case if we use the canonical US Standard model for the atmosphere which is not accurate enough for the level of precision we demand.

We are currently working on providing the composition of ultra-high energy cosmic rays using the radio signal.


\end{document}